# Reconstruction of Power System Measurements Based on Enhanced Denoising Autoencoder

You Lin
Department of Electrical Engineering
Southern Methodist University
Dallas, TX, USA
youl@smu.edu

Jianhui Wang
Department of Electrical Engineering
Southern Methodist University
Dallas, TX, USA
jianhui@smu.edu

Mingjian Cui
Department of Electrical Engineering
Southern Methodist University
Dallas, TX, USA
mingjiancui@smu.edu

*Abstract*—This paper presents a new solution for reconstructing missing data in power system measurements. An Enhanced Denoising Autoencoder (EDAE) is proposed to reconstruct the missing data through the input vector space reconstruction based on the neighbor values correlation and Long Short-Term Memory (LSTM) networks. The proposed LSTM-EDAE is able to remove the noise, extract principle features of the dataset, and reconstruct the missing information for new inputs. The paper shows that the utilization of neighbor correlation can perform better in missing data reconstruction. Trained with LSTM networks, the EDAE is more effective in coping with big data in power systems and obtains a better performance than the neural network in conventional Denoising Autoencoder. A random data sequence and the simulated Phasor Measurement Unit (PMU) data of power system are utilized to verify the effectiveness of the proposed LSTM-EDAE.

## I. Introduction

With the increasing development of techniques in data measurement, transmission, and analysis in modern power systems, measurements are widely used in applications of power system state estimation, fault diagnosis, load and renewable generation prediction, and so on [1-3]. Most of the state-of-the-art researches are based on the assumption that power system measurements are ideal and not tampered. However, in fact, measurements are frequently corrupted in the processes of production, collection, transferring, and analysis, such as data entry errors caused by human behaviors, measuring errors caused by the communication device outage, experimental errors caused by the data sampling or extraction [4]. Hence, it is essential to reconstruct a corrupted dataset to provide a more precise data condition for the existing techniques.

Numerous methodologies have been done in data reconstruction, including the regression method [5, 6], the interpolation method [7, 8], and the data compression and reconstruction methods [9, 10]. The regression model adopts corresponding historical measurements as inputs to fit the interested missing values. General regression methods can be utilized, such as support vector machine, neural networks, and kernel density estimation methods. A nonlinear regression model for geomagnetic data reconstruction is constructed in [5] by combining different machine learning methods using the historical data as input and the missing data as the desired output. Noise suppression methods based on the partial differential equation are proposed in [11] to process signals with a smooth change in the intensity value, which is usually utilized in image processing with continuous variation. The interpolation methods are usually based on the weighted summation of the neighbor values of the missing data. In [12], the weighted average values in the neighborhood are used to replace the missing data in the seismic data reconstruction, in which the normalized Gaussian weighted filter is utilized. Sparsity model based on the data compression or feature extraction is demonstrated to be an effective way to extract typical characteristics from the corrupted data [13, 14]. The data compression process will enable us to feature principle characteristics of the dataset with a sparsity model. The undersampled data can be reconstructed consistent with the acquired samples and compliant with a sparsity model in some transform domains.

Autoencoders (AEs) are widely used for data compression and reconstruction because of its high nonlinearity. The trained AEs are utilized to recompose the missing information provided by the SCADA measurements in power systems [13], building energy data [15], and the contaminated images [10]. However, when the number of applied nodes in the hidden layer of neural networks is more than the dimension of inputs, the network may learn the identity function, which may make the AE learn nothing useful [16]. That is, if a corrupted input is given, the AE could learn the corrupted dataset and consequently generate the same output. Denoising AEs (DAEs) can avoid this problem by corrupting the data on purpose through randomly turning some input values to zero [17]. However, the DAE can only generate a fixed value for all inputs which are corrupted by zeros. That is, for the trained DAE, all the missing values in the new input vectors will be reconstructed by a same value. Therefore, the trained DAE will fail to recognize the varying features of the missing values. Moreover, the hidden layers of classic DAEs are based on conventional neural networks. In the modern power system, a huge number of measurements are collected, conventional neural networks in DAE are not applicable any more. The development of deep learning method enables us to cope with big data in real practice.

Based on the above analysis in this paper, we propose an enhanced deep learning-based DAE considering the varying features of missing data and an advanced deep learning method based on Long Short-Term Memory (LSTM) networks. Section II describes the basics and problems of conventional DAE with examples. Section III presents the proposed LSTM-EDAE method in detail. In Section III, the varying features of the missing data are represented by their neighbor values and the input vector space of DAE is expanded through vector space reconstruction. Then, hidden layers based on LSTM in the proposed EDAE are described. Case studies based on the random sequence and simulated PMU measurements are presented in Section IV. Section V concludes the paper.

## II. DENOISING AUTOENCODER

### A. Basics of Autoencoder

AE is an enhanced Principal Component Analysis (PCA) method. PCA is a linear model for feature extraction. While AE has a nonlinear transformation unit, the learned model may be more refined and have a stronger expression of inputs. AE is a type of neural networks (NNs) that aims to copy inputs to outputs [18]. It works by compressing the input into a latent-space representation, and then reconstructing the output from this representation.

AE is an unsupervised training model including the encoder and decoder process, as shown in Fig. 1. An input vector $x$ is defined. In the encoder process, the input $x$ is mapped to a compressed representation $y$ through the mapping function $y = f_\theta(x) = s(Wx + b)$ [18]. Here the parameter set $\theta = \{W, b\}$ is the collection of weights $W$ and bias $b$ to be estimated in the training process. In the decoder process, the compressed representation $y$ is mapped back to a vector $z$ for reconstructing the input $x$ with the mapping function $z = g_{\theta'}(y) = s(W'y + b')$ with the parameter set $\theta' = \{W', b'\}$. With $N$ input vectors $\{x_i\}_{i=1}^N$, the optimal solution of parameters $\theta$ and $\theta'$ can be estimated by minimizing the average reconstruction error represented by the loss function $J(x, z)$, given by:

$$\{\theta_{opt}, \theta'_{opt}\} = \arg\min_{\theta, \theta'} \frac{1}{N} \sum_{n=1}^{N} J(x_i, z_i) \quad (1)$$

where the loss function $J$ can be the squared error of the input $x_i$ and its corresponding reconstruction $z_i$; $\theta_{opt} = \{W_{opt}, b_{opt}\}$ and $\theta'_{opt} = \{W'_{opt}, b'_{opt}\}$. In order to improve the robustness of the AE, a sparsity penalty term is added in the objective function in (1). In the training process, the trained AE is obtained with optimal parameters $\theta_{opt}$ and $\theta'_{opt}$.

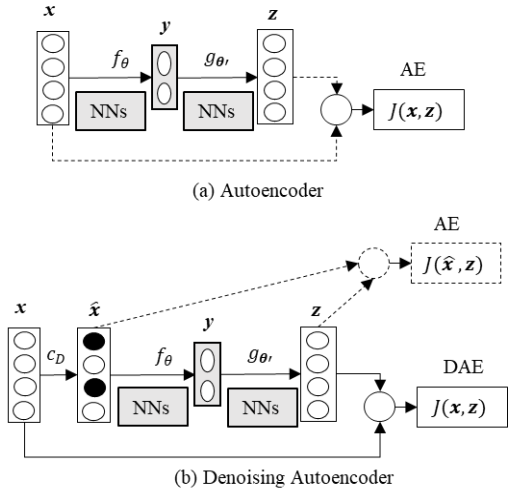

Fig. 1. Autoencoder and Denoising Autoencoder.

### B. Pros and Cons of DAE

If the initial input $x$ is corrupted by random zeros, the conventional AE will learn and reconstruct the corrupted input vector $\hat{x}$ in the training process, which is useless in the real application. To intuitively show this problem, we give a simple example in Example 1.

***Example 1***: Given a corrupted input with one-dimensional dataset $X = \{x_i\}_{i=1}^{1000}$, which is marked with red dots in Fig. 2. In this figure, the corrupted values in $X$ which are being set to 0 are marked with green circles. The reconstruction results of the dataset $X$ using the trained AE are marked with black dots. The corresponding reconstruction results of the corrupted values in $X$ are marked with yellow circles. The reconstructed results are almost similar with the corrupted dataset without an appropriate estimation of the missing values or the corrupted values.

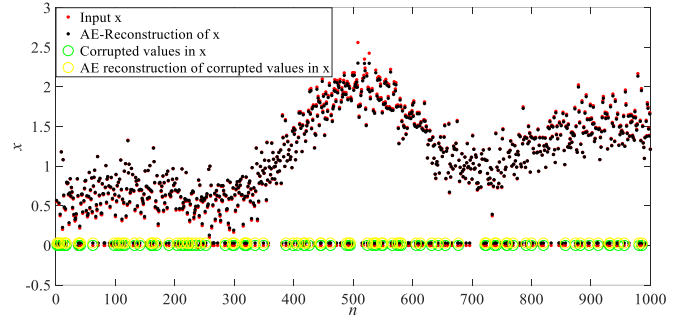

Fig. 2. Results of Example 1: data reconstruction from Autoencoder.

DAE solves the above problem through modifying the loss function in the training process. In DAE, the corrupted input $\hat{x}$ is simulated by a partially destroyed version of $x$, as shown in (2). The stochastic mapping $c_D$ randomly turns some of the elements in $x$ to 0.

$$\hat{x} = c_D(x) \quad (2)$$

The DAE learns to remove the zero noise by modifying the loss function with comparing the initial input $x$ and the reconstruction results $z$ which is the reconstruction of the corrupted input $\hat{x}$ [17]. The loss function here is $J(x, z)$ instead of $J(\hat{x}, z)$, as shown in Fig. 1. Through this modification, the DAE is able to learn something useful and remove noises in $\hat{x}$.

The improvement of DAE is data compression and reconstruction of the corrupted input $\hat{x}$. It outputs a relatively better value than zero at the corrupted points. However, with the trained DAE, one can get the same data reconstruction value for the missing values when replacing the missing value with zero. The reason is that the conventional DAE only takes values of $\hat{x}$ as inputs without considering its varying characteristics of neighbor values. With the same input, the trained DAE outputs the same value. An example of one-dimension data is tested in Example 2 to give an intuitive explanation of this problem.

***Example 2***: For one-dimension data, with the same input $\hat{x} = 0$, the trained DAE gives the same reconstructed value, as shown in Fig. 3. The reconstruction results of the corrupted values are shifted up to a fixed value instead of zero to minimize the modified loss function. As can be seen, the reconstruction results are not optimal.

Thus, we solve this problem in the proposed Enhanced Denoising Autoencoder (EDAE) by considering the neighbor values of the missing data.

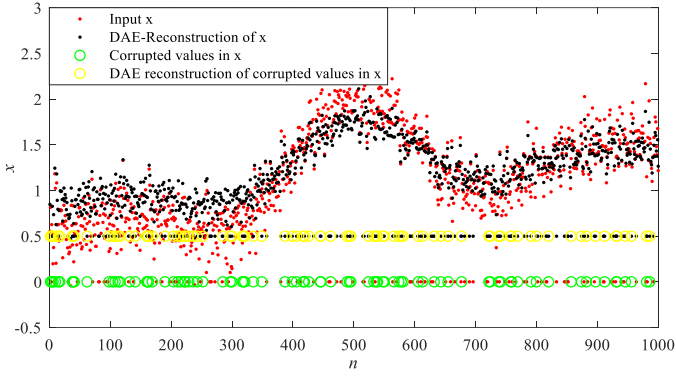

Fig. 3. Results of Example 2: data reconstruction from Denoising Autoencoder

## III. ENHANCED DEEP LEARNING BASED DENOISING AUTOENCODER

To deal with the above problems, an EDAE is proposed in this paper. The main idea is to supplement the missing data in the new input dataset with fake values. The fake values should have similar features as the good data. Thus, we take advantage of the geographical neighbors to mimic the fake values of the missing data. In addition, the hidden layers of classic DAE are based on neural networks (NNs). In the modern power system, due to a large number of measurements collected, it is essential to find deep learning algorithms adaptive for big data in power systems [1]. NNs in the conventional DAE are not applicable any more. Thus, we adopt the LSTM networks which works better in dealing with huge amount of data when enough training data is available.

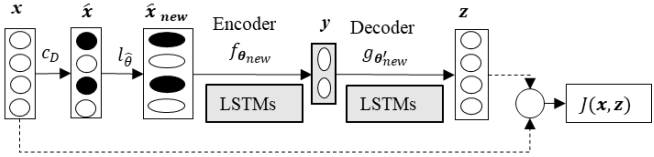

Fig. 4. Architecture of EDAE.

### A. Initialize the Missing Values in the Corrupted Dataset based on Geographical Similarity

'Geography' here means the location of one value in a time sequence. Neighbor values in the sequence can be taken as 'Geography neighbors'. When a time sequence has good autocorrelation features, the neighbor samples in the sequence will have good correlation, which is called geographical similarity. Thus, it is possible to take advantage of the geographical priorities in the time sequence, which is in accordance with the fact of power system measurements. The geographical similarity is also utilized in the reconstruction methods based on regression models [7]. However, in these methods, the missing data is simulated based on the neighbor values without considering the overall features of the dataset. In this paper, fake data is only simulated to replace the missing data in the input vector.

Suppose the missing data at time $t$ is represented by $x_t$. Based on the correlation theory, it is possible to estimate a fake value from its past $K_b$ and future $K_f$ neighbor vectors, given by:

$$x_t = l_{\hat{\theta}}(x_{t-K_b}, x_{t-K_b+1}, \cdots, x_{t-1}, x_{t+1}, \cdots, x_{t+K_f+1}, x_{t+K_f}) \quad (3)$$

where, $l$ is the fit function between the $K_b + K_f$ neighbor vectors and the missing data $x_t$; $\hat{\theta}$ is the implicit parameters to be estimated. Thus, in the EDAE, the mapping process from $\hat{x}$ to the compressed representation $y$ can be formulated as:

$$\begin{aligned} y_t &= f_\theta(x) \\ &= f_\theta(l_{\hat{\theta}}(x_{t-K_b}, x_{t-K_b+1}, \cdots, x_{t-1}, x_{t+1}, \cdots, x_{t+K_f+1}, x_{t+K_f})) \\ &= s(W_t l_{\hat{\theta}}(x_{t-K_b}, x_{t-K_b+1}, \cdots, x_{t-1}, x_{t+1}, \cdots, x_{t+K_f+1}, x_{t+K_f}) + b_t) \end{aligned} \quad (4)$$

The above mapping function can be modified to the simplified expression in (5) if we reconstruct the vector space by expanding the original vector space in $x_t$ which is a $L \times 1$ vector to a new vector space as $x_{new}$ as shown in (6) whose dimension is $(K_b + K_f + 1)L \times 1$. The mapping process is shown in Fig. 4.

$$y_t = f_{\theta_{new}}(x_{new}) = s(W_{t,new} x_{new} + b_{t,new}) \quad (5)$$

$$x_{new} = [x_{t-K_b}^T, x_{t-K_b+1}^T, \cdots, x_{t-1}^T, x_t^T, x_{t+1}^T, \cdots, x_{t+K_f-1}^T, x_{t+K_f}^T]^T \quad (6)$$

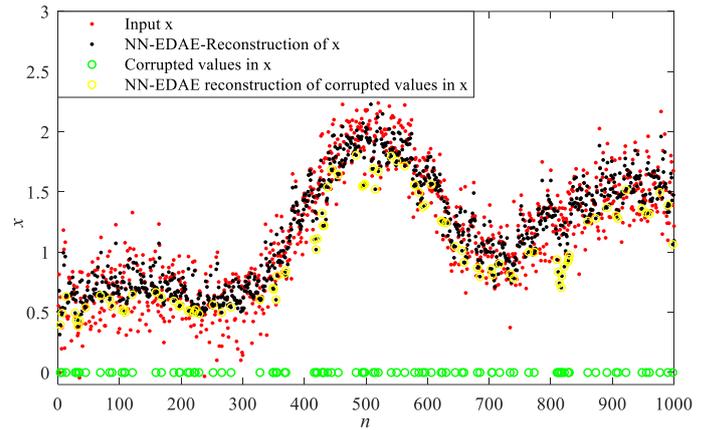

Fig. 5. Results of Example 3: data reconstruction from NN-EDAE.

Therefore, based on the vector space reconstruction, the optimal parameters in the encoder and decoder process can be trained using advanced deep learning algorithms. To simply demonstrate its effectiveness, we present a test for the reconstruction of a one-dimension corrupted data, as shown in Example 3.

*Example 3*: The data to be reconstructed is similar to that in Examples 1 and 2. $K_b = K_f = 5$ is utilized to reconstruct the input vector space. The results are shown in Fig. 5. As can be seen, the reconstructed values marked with yellow circles of the corrupted values marked with green circles are much more accurate than the results in Figs. 2 and 3.

### B. LTSM-based Enhanced Denoising Autoencoder

LTSM network is a special kind of recurrent neural network which is capable of taking advantage of previous time steps and learning long-term information. It was firstly introduced by Hochreiter & Schmidhuber in 1997 [19]. Thus, LSTM network is utilized to replace the neural network in convential DAE, as shown in Fig. 4, to make it more powerful in encoder and decoder functions.

The LSTM contains memory units in the recurrent hidden layer. LSTM neural is able to restore temporal state of the network through self-connections. Meanwhile, information flowing through the memory unit can be controlled with input gate $i$, output gate $o$, forget gate $f$ and the memory unit activation vector $c$. The improvements in LSTM enable the neural to read, store, forget or reset the memory unit. Given the input sequence $\{x_n\}_{n=1}^{N}$, the output sequence $\{y_n\}_{n=1}^{N}$ can be obtained by the following equations representing the network unit:

$$\begin{aligned}
i_n &= \sigma(W_{ix}x_n + W_{im}m_{n-1} + W_{ic}c_{n-1} + b_i) \\
f_n &= \sigma(W_{fx}x_n + W_{fm}m_{n-1} + W_{fc}c_{n-1} + b_f) \\
c_n &= f_n \odot c_{n-1} + i_n \odot g(W_{cx}x_n + W_{cm}m_{n-1} + b_c) \\
o_n &= \sigma(W_{ox}x_n + W_{om}m_{n-1} + W_{oc}c_n + b_o) \\
m_n &= o_n \odot h(c_n) \\
y_n &= \phi(W_{ym}m_n + b_y)
\end{aligned} \quad (7)$$

where, $W$ terms represent the weight matrices, $b$ terms represent bias vectors; $h$, $g$ and $\phi$ are activation functions; $\odot$ represents the element-wise product of the vectors.

## IV. CASE STUDY

To intuitively show the universal effectiveness of the proposed LSTM-EDAE method for data reconstruction, we perform two case studies in this section. Case A is based on a random data sequence. Case B is based on simulated PMU measurements in power systems. The proposed LSTM-EDAE approach is utilized to reconstruct corrupted data to verify its effeteness. The AE, DAE, the regression method based on the Extreme Learning Machine (ELM) and the Interpolation method (IM) based on the average of neighbor values are utilized as benchmarks to make comparisons with the proposed LSTM-EDAE.

### A. Case A: An Example Tests on Random Sequences

One-dimension data sequence $X = \{x_i\}_{i=1}^{1000}$ is generated by (8):

$$\begin{cases} x(n) = 1 + r(n) \cdot 5e^{-2} + \dfrac{\sin(r(n))}{r(n)} + 0.4 \cdot \varepsilon_1(n) \\ r(n) = -10 + 20 \cdot \varepsilon_2(n) \end{cases}, n = 1, \cdots, 1000 \quad (8)$$

where, $\varepsilon_1(n)$ and $\varepsilon_2(n)$ are random Gaussian noises with zero mean and unit variance.

In this case, data sequence $X$ is corrupted with zeros in different proportions $\rho$ = 10%, 20%, 30%, 40%, and 50%. The reconstruction results using AE, DAE, and the proposed LSTM-EDAE corresponding to $\rho$ = 20% are shown in Fig. 6, respectively. This figure intuitively demonstrates that the proposed LSTM-EDAE can obtain much better reconstructions of the corrupted values in $X$ even for the corrupted proportion 20%. The normalized mean square errors (NMSEs) of the reconstruction results with zero corruptions in different proportions are shown in Table I. The results show that reconstruction errors of the proposed LSTM-EDAE are relatively small even when 50% data are missing. In addition, the performance of the proposed method is less sensitive to the increase of the corrupted proportion.

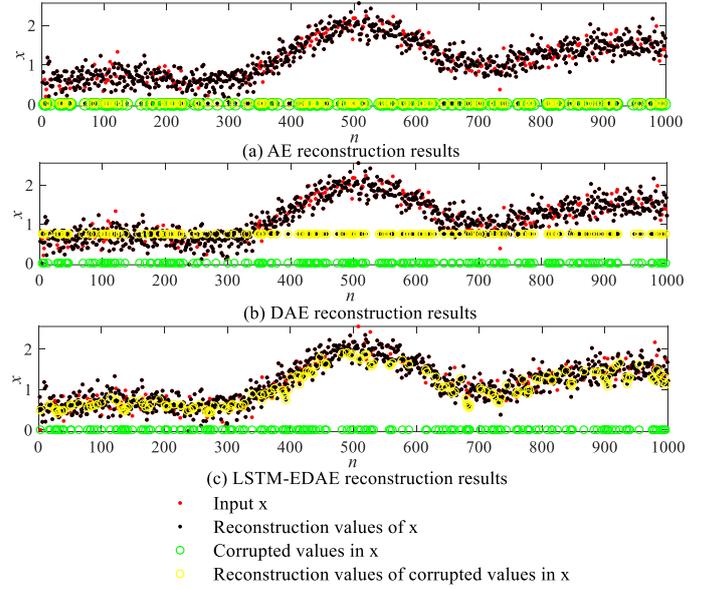

- • Input x
- • Reconstruction values of x
- ○ Corrupted values in x
- ○ Reconstruction values of corrupted values in x

Fig. 6. Reconstruction results based on AE, DAE, and the proposed LSTM-EDAE. ($\rho$ = 20%)

TABLE I. NMSEs OF THE RECONSTRUCTION RESULTS WITH ZERO CORRUPTIONS IN DIFFERENT PROPORTIONS

| Methods | Corrupted proportions $\rho$ | | | | |
|---|---|---|---|---|---|
| | $\rho$ = 10% | $\rho$ = 20% | $\rho$ = 30% | $\rho$ = 40% | $\rho$ = 50% |
| AE | 0.0217 | 0.0422 | 0.0672 | 0.0981 | 0.1100 |
| DAE | 0.0099 | 0.0116 | 0.0148 | 0.0178 | 0.0198 |
| LSTM-EDAE | 0.0011 | 0.0024 | 0.0069 | 0.0087 | 0.0116 |

### B. Case B: Tests on Simulated PMU Measurments in Power Systems

Simulated PMU measurements of feeders in power systems are utilized to verify the effectiveness of the proposed LSTM-EDAE. The simulated PMU measurements are based on the structure in [20]. The time resolution is 1 minute. The tested variable is three phase real power which is calculated from the measured three phase voltages and currents. We simulate the missing data in real power series by replacing a specific portion of all samples with zero. The reconstruction results of real power $P_a$ in phase A of one feeder with 10% corruptions using AE, DAE, and the proposed LSTM-EDAE are shown in Fig. 7. It is found that AE fails to reconstruct the corrupted values. The proposed LSTM-EDAE method shows a better performance than the conventional DAE. As shown in Fig. 7, the reconstruction results of the proposed LSTM-EDAE are close to the initial input value. The NMSEs of the reconstruction results with zero corruptions in different proportions are shown in Table II. Table II indicates that the IM has good accuracy similar as the proposed LSTM-EDAE method when 10% values are missing. However, when the proportion of missing values increase, the reconstruction error of the proposed LSTM-EDAE are much smaller than other benchmarks. Reconstruction results using IM, ELM, and the proposed LSTM-EDAE with 30% values missing are shown in Fig. 8. The figure demonstrates that the proposed LSTM-EDAE has better accuracy than other benchmarks in data reconstruction.

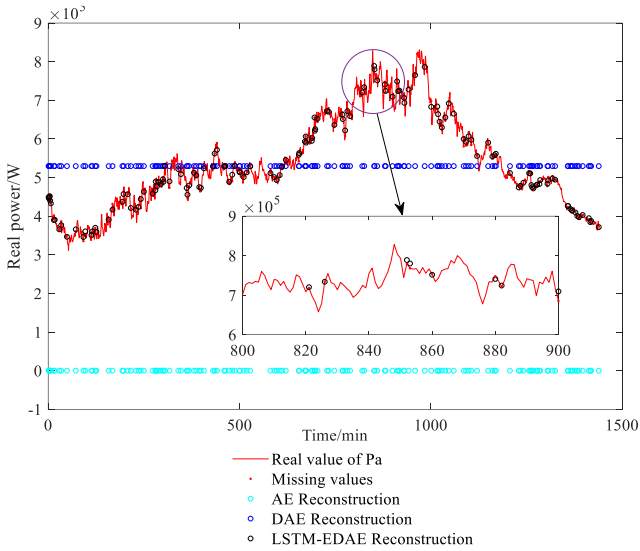

Fig. 7. Reconstruction results using AE, DAE, and the proposed LSTM-EDAE. ($\rho = 10\%$)

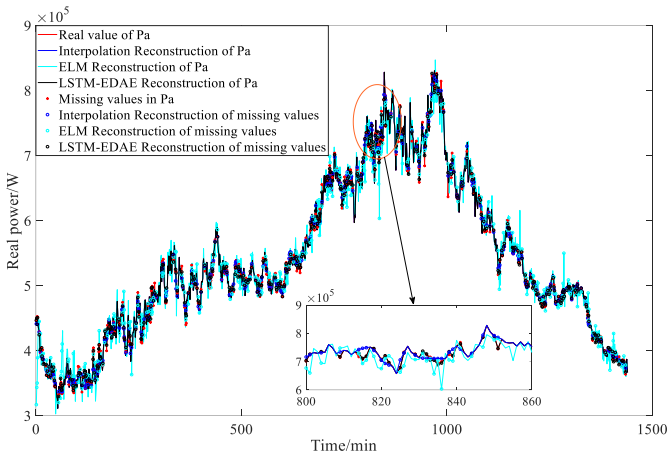

Fig. 8. Reconstruction results using IM, ELM, and the proposed LSTM-EDAE. ($\rho = 30\%$)

TABLE II. NMSEs of the Reconstruction Results with Zero Corruptions in Different Proportions

| Methods | Corrupted proportions $\rho$ | | | | |
|---|---|---|---|---|---|
| | $\rho = 10\%$ | $\rho = 20\%$ | $\rho = 30\%$ | $\rho = 40\%$ | $\rho = 50\%$ |
| AE | 0.0173 | 0.0414 | 0.0693 | 0.0867 | 0.1049 |
| DAE | 0.0088 | 0.0151 | 0.0193 | 0.0249 | 0.0289 |
| ELM | 0.0009 | 0.0033 | 0.0045 | 0.0074 | 0.0104 |
| IM | 0.0007 | 0.0022 | 0.0041 | 0.0070 | 0.0101 |
| LSTM-EDAE | 0.0006 | 0.0020 | 0.0034 | 0.0062 | 0.0085 |

## V. Conclusion

A novel LSTM-based Enhanced Denoising Autoencoder (LSTM-EDAE) is proposed to perform the data reconstruction based on the vector space reconstruction of input vectors. The test results of the power system measurements are in accordance with those of random sequences, which demonstrates that the proposed LSTM-EDAE method has good robustness and universality in applications. The test results verify that the proposed EDAE has a better performance than the conventional Autoencoder (AE) and Denoising Autoencoder (DAE). Uncertain features of the missing data in data reconstruction will be considered in our future study.